\documentclass[aps,prb,reprint,amssymb, amsmath, superscriptaddress, longbibliography,floatfix]{revtex4-2}

\usepackage{lipsum, xcolor,graphicx, amsmath}
\usepackage{placeins, float}
\usepackage{amssymb}
\usepackage{comment}
\usepackage{braket}
\usepackage{balance}
\usepackage[dvipsnames]{xcolor}
\newcommand{\sr}[1]{\textcolor{black}{#1}}
\DeclareUnicodeCharacter{2062}{}
\begin{document}
%%%%%%%%%%%%%%%%%%%%%%%%%%%%%%%%%%%%%%%%%%%%%%%%%%%%%%%%%%%%%%
%%%%%%%%%%%%%%%%%%%%%%%%%%%%%%%%%%%%%%%%%%%%%%%%%%%%%%%%%%%%%%
%%%%%%%%%%%%%%%%%%%%%%%%%%% TITLE %%%%%%%%%%%%%%%%%%%%%%%%%%%%
%%%%%%%%%%%%%%%%%%%%%%%%%%%%%%%%%%%%%%%%%%%%%%%%%%%%%%%%%%%%%%
%%%%%%%%%%%%%%%%%%%%%%%%%%%%%%%%%%%%%%%%%%%%%%%%%%%%%%%%%%%%%%

\title{Trigonal distortion in the Kitaev candidate honeycomb magnet BaCo$_{2}$(AsO$_{4})_{2}$ }
%%%%%%%%%%%%%%%%%%%%%%%%%%%%%%%%%%%%%%%%%%%%%%%%%%%%%%%%%%%%%%
%%%%%%%%%%%%%%%%%%%%%%%%%%%%%%%%%%%%%%%%%%%%%%%%%%%%%%%%%%%%%%
%%%%%%%%%%%%%%%%%%%%%%%% AUTHOR LIST %%%%%%%%%%%%%%%%%%%%%%%%%
%%%%%%%%%%%%%%%%%%%%%%%%%%%%%%%%%%%%%%%%%%%%%%%%%%%%%%%%%%%%%%
%%%%%%%%%%%%%%%%%%%%%%%%%%%%%%%%%%%%%%%%%%%%%%%%%%%%%%%%%%%%%%

\author{M.~M.~Ferreira-Carvalho}
\affiliation{Max Planck Institute for Chemical Physics of Solids, N{\"o}thnitzer Str. 40, 01187 Dresden, Germany}
\affiliation{Institute of Physics II, University of Cologne, Zülpicher Straße 77, 50937 Cologne, Germany }

\author{S.~R\"o{\ss}ler}
\affiliation{Max Planck Institute for Chemical Physics of Solids, N{\"o}thnitzer Str. 40, 01187 Dresden, Germany}

\author{C.~F.~Chang}
\affiliation{Max Planck Institute for Chemical Physics of Solids, N{\"o}thnitzer Str. 40, 01187 Dresden, Germany}

\author{Z.~Hu}
\affiliation{Max Planck Institute for Chemical Physics of Solids, N{\"o}thnitzer Str. 40, 01187 Dresden, Germany}

\author{S.~M.~Valvidares}
\affiliation{ALBA Synchrotron Light Source, E-08290 Cerdanyola del Vall\`{e}s, Barcelona, Spain}

\author{P.~Gargiani}
\affiliation{ALBA Synchrotron Light Source, E-08290 Cerdanyola del Vall\`{e}s, Barcelona, Spain}

\author{M.~W.~Haverkort}
\affiliation{Institute for theoretical physics, Heidelberg University, Philosophenweg 19, 69120 Heidelberg, Germany}

\author{Prashanta K. Mukharjee}
\affiliation{Experimental Physics VI, center for Electronic Correlations and Magnetism, Institute of Physics, University of Augsburg, 86159 Augsburg, Germany}

\author{P.~Gegenwart}
\affiliation{Experimental Physics VI, Center for Electronic Correlations and Magnetism, Institute of Physics, University of Augsburg, 86159 Augsburg, Germany}

\author{A.~A.~Tsirlin}
\affiliation{Felix Bloch Institute for Solid-State Physics, University of Leipzig, 04103 Leipzig, Germany}

\author{L.~H.~Tjeng}
\affiliation{Max Planck Institute for Chemical Physics of Solids, N{\"o}thnitzer Str. 40, 01187 Dresden, Germany}

\date{\today}
%%%%%%%%%%%%%%%%%%%%%%%%%%%%%%%%%%%%%%%%%%%%%%%%%%%%%%%%%%%%%%
%%%%%%%%%%%%%%%%%%%%%%%%%%%%%%%%%%%%%%%%%%%%%%%%%%%%%%%%%%%%%%
%%%%%%%%%%%%%%%%%%%%%%%%%% ABSTRACT %%%%%%%%%%%%%%%%%%%%%%%%%%
%%%%%%%%%%%%%%%%%%%%%%%%%%%%%%%%%%%%%%%%%%%%%%%%%%%%%%%%%%%%%%
%%%%%%%%%%%%%%%%%%%%%%%%%%%%%%%%%%%%%%%%%%%%%%%%%%%%%%%%%%%%%%

\begin{abstract}
We conducted x-ray absorption (XAS) and magnetic circular dichroism (XMCD) measurements at the Co $L_{2,3}$ edges on single crystals of the Kitaev candidate honeycomb lattice compound BaCo$_2$(AsO$_4$)$_2$. The measurements employed the inverse partial fluorescence yield technique, which is ideal for acquiring reliable x-ray absorption spectra from highly insulating samples, enabling precise quantitative analysis. Our experimental results revealed a significant linear dichroic signal, indicating strong trigonal distortion in the CoO$_{6}$ octahedra in BaCo$_2$(AsO$_4$)$_2$. We performed a detailed analysis of the experimental XAS and XMCD spectra using a full-multiplet configuration-interaction cluster model. This analysis unveiled that the $t_{2g}$ hole density is predominantly localized in the $a_{1g}$ orbital. Through XMCD sum rules and theoretical calculations, we quantified both the spin and orbital magnetic moments. Our study demonstrates that the local electronic structure of the CoO$_{6}$ octahedra displays an effective trigonal distortion of approximately $-0.114$ eV. This distortion is larger than the Co $3d$ spin-orbit coupling constant, emphasizing the crucial impact of local structural distortions on the electronic and magnetic properties of BaCo$_2$(AsO$_4$)$_2$.
\end{abstract}

\maketitle
%%%%%%%%%%%%%%%%%%%%%%%%%%%%%%%%%%%%%%%%%%%%%%%%%%%%%%%%%%%%%%
%%%%%%%%%%%%%%%%%%%%%%%%%%%%%%%%%%%%%%%%%%%%%%%%%%%%%%%%%%%%%%
%%%%%%%%%%%%%%%%%%%%%%%% INTRODUCTION %%%%%%%%%%%%%%%%%%%%%%%%
%%%%%%%%%%%%%%%%%%%%%%%%%%%%%%%%%%%%%%%%%%%%%%%%%%%%%%%%%%%%%%
%%%%%%%%%%%%%%%%%%%%%%%%%%%%%%%%%%%%%%%%%%%%%%%%%%%%%%%%%%%%%%

\section{Introduction}
%\subsection{}
%\subsubsection{}
Novel and exotic quantum phases of matter display a plethora of physics that conventional mean field theories may not capture. One of such states is the quantum spin liquid (QSL) groundstate which does not exhibit long-range magnetic order but is characterized by topological order with fractionalized excitations and long-range entanglement,  displaying highly correlated spins that fluctuate even at absolute zero temperature \cite{Broh2020}. Such systems were initially explored in geometrically frustrated magnetic structures with $S = 1/2$, but recently the search for the materialization of the QSL groundstate became tremendously intensified with the theoretical implementation of the exactly solvable Kitaev model \cite{Kitaev2006} for $J_\mathrm{eff}\,=\,1/2$ pseudospin systems with the honeycomb lattice structure \cite{Jack2009}. In this model, neighboring spins are coupled in a highly anisotropic manner with bond-dependent interactions originating from the spin-orbit coupling, which give rise to localized and itinerant Majorana fermions. 

\begin{figure}[t]
	\centering
	\includegraphics[width=\linewidth]{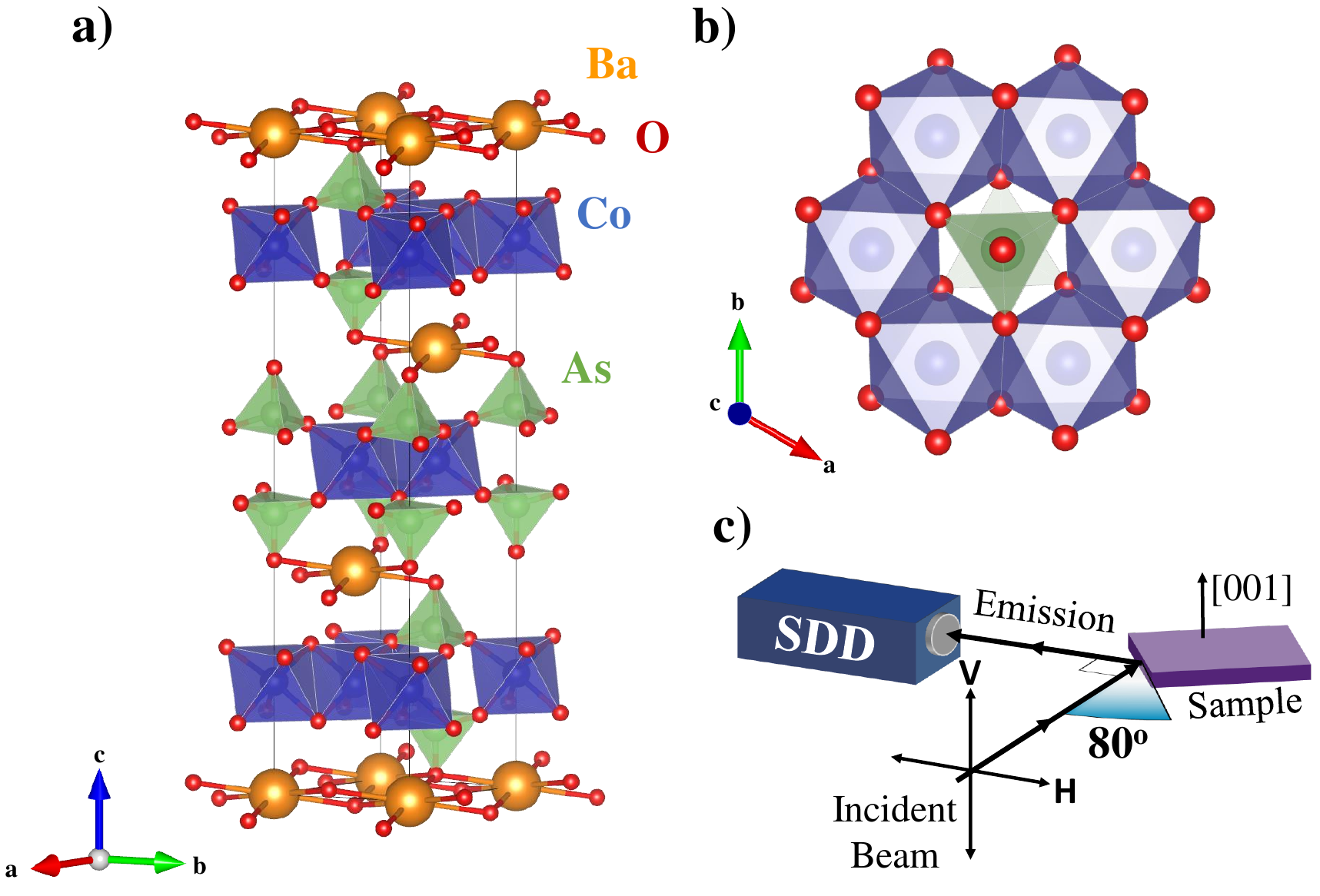}
	\caption{(a) Crystal structure of BaCo$_2$(AsO$_4$)$_2$ displaying layered arrangement of CoO$_{6}$ octahedra along the $c-$axis. (b) The $ab$ plane containing honeycomb lattice of edge-sharing CoO$_{6}$ octahedra with AsO$_{4}$ tetrahedra at the center. (c) The experimental geometry of the incident beam, sample, and the SDD detector.}
	\label{fig:geometry}
\end{figure}

Initially, most studies were focused on compounds containing $4d$ and $5d$ transition metal ions (i.e., Ru$^{3+}$, Rh$^{4+}$, and Ir$^{4+}$) in edge-sharing octahedral geometry \cite{Win2017,Takagi2019,Tre2022}. However, it was found that the real-world honeycomb materials always feature sizable non-Kitaev interactions \cite{Chal2010} that lead to magnetic order at low temperatures. Yet, the expectation remains that such materials may be tuned into the QSL phase via external parameters such as pressure, uniaxial strain, or applied magnetic field. In $\alpha$-RuCl$_{3}$, for example,
the zig-zag antiferromagnetic order which appears at $T_N\,\approx$ 7-13\,K could be suppressed by the application of an in-plane magnetic field of $\approx\,8\,$T \cite{John2015,Ben2018}. This field-induced phase was quickly characterized as the Kitaev QSL phase \cite{Baek2017}. Nevertheless, this assessment is currently still subject of intense debate \cite{Tre2022}.

Very recently, it has been proposed that honeycomb lattice based compounds with Co$^{2+}$ ions in octahedral coordination could be suitable candidates for the realization of a QSL groundstate \cite{Liu2018,Kim2022}. In the presence of the octahedral crystal field, the high-spin $3d^{7}$: $t_{2g}^{5}e_{g}^{2}$ ($S\,=\,3/2$, $L\,=\,1$) configuration also posses the effective $J_\mathrm{eff}\,=\,1/2$ groundstate due to the spin-orbit coupling (SOC), very similar to the $4d^5$ and $5d^5$ Kitaev candidate materials. Moreover, the presence of the $e_{g}$ electrons in the ground state configuration leads to more exchange channels which should further suppress the Heisenberg and off-diagonal exchange interactions due to their opposite sign\,\cite{Liu2020}. Thus, Co$^{2+}$ honeycomb materials seem to provide improved conditions for the realization of the Kitaev QSL groundstate. However, details of the actual crystal structure of the material may matter since the presence of effective crystal fields of lower than octahedral symmetry could cause serious deviations from the ideal $J_\mathrm{eff}\,=\,1/2$ state \cite{Win2022,Liu2023}. This is the focus of our study on the candidate material BaCo$_2$(AsO$_4$)$_2$ in which we determine the local electronic and magnetic properties of the Co$^{2+}$ ions using linearly polarized x-ray absorption spectroscopy (XAS) and x-ray magnetic circular dichroims (XMCD) at the $L_{2,3}$ edges. 

BaCo$_2$(AsO$_4$)$_2$ shows several intriguing properties, most notably, the suppression of long-range magnetic order by in-plane magnetic field of $\approx\,0.5$\,T \cite{Zho2020}. It crystalizes in the hexagonal $R\bar{3}$ space group, consisting of layers of edge-sharing Co$^{2+}$ octahedra which are sandwiched in between AsO$_4$ tetrahedra along the $c-$axis as depicted in Fig.\,1\,(a). The edge-sharing CoO$_{6}$ octahedra forming the honeycomb lattice in the $ab$ plane is shown in Fig.\,1\,(b). In this structure, the  Co$^{2+}$ ions have a site symmetry $C_{3}$. The material orders antiferromagnetically at a temperature $T_N\,\approx 5.4 $\,K. The suppression of magnetic order by a small magnetic field suggests a possible field-induced Kitaev QSL groundstate \cite{Zho2020}. A finite residual linear term of thermal conductivity observed in this field regime has been attributed to mobile fractionalized spinon excitation \cite{Tu2024}.

On the other hand, magnon excitations found in a time domain terahertz spectroscopy measurements \cite{Shi2021} have indicated that the field-induced state in the field regime of 0.5\,T, is not compatible with a Kitaev QSL. Furthermore, a recent study consisting of magnetometry, calorimetry, high-resolution capacities dilatometry, and single crystal neutron diffraction experiments reported that a magnetic Bragg peak persists in the proposed QSL phase with applied in-plane magnetic fields \cite{Mukh2024}.  However, the exact microscopic model of this material, the
size and even the sign of the Kitaev term remain vividly debated \cite{Lee2024,Lee2025,Safari2024,Maksimov2025,Devillez2025}. In fact, one would readily expect that a trigonal distortion in these systems could modify the psuedospin wave function, affecting the non-Kitaev terms \cite{Liu2020,Liu2023,Sam2024} and thus strongly influencing the $J/K$ ratio. Hence it is vitally important to determine the magnitude of the trigonal distortion since this gives an estimate of the energy-scale required for tuning the system into the QSL regime. 

\begin{figure*}[ht]
	\centering
	\includegraphics[width=\linewidth]{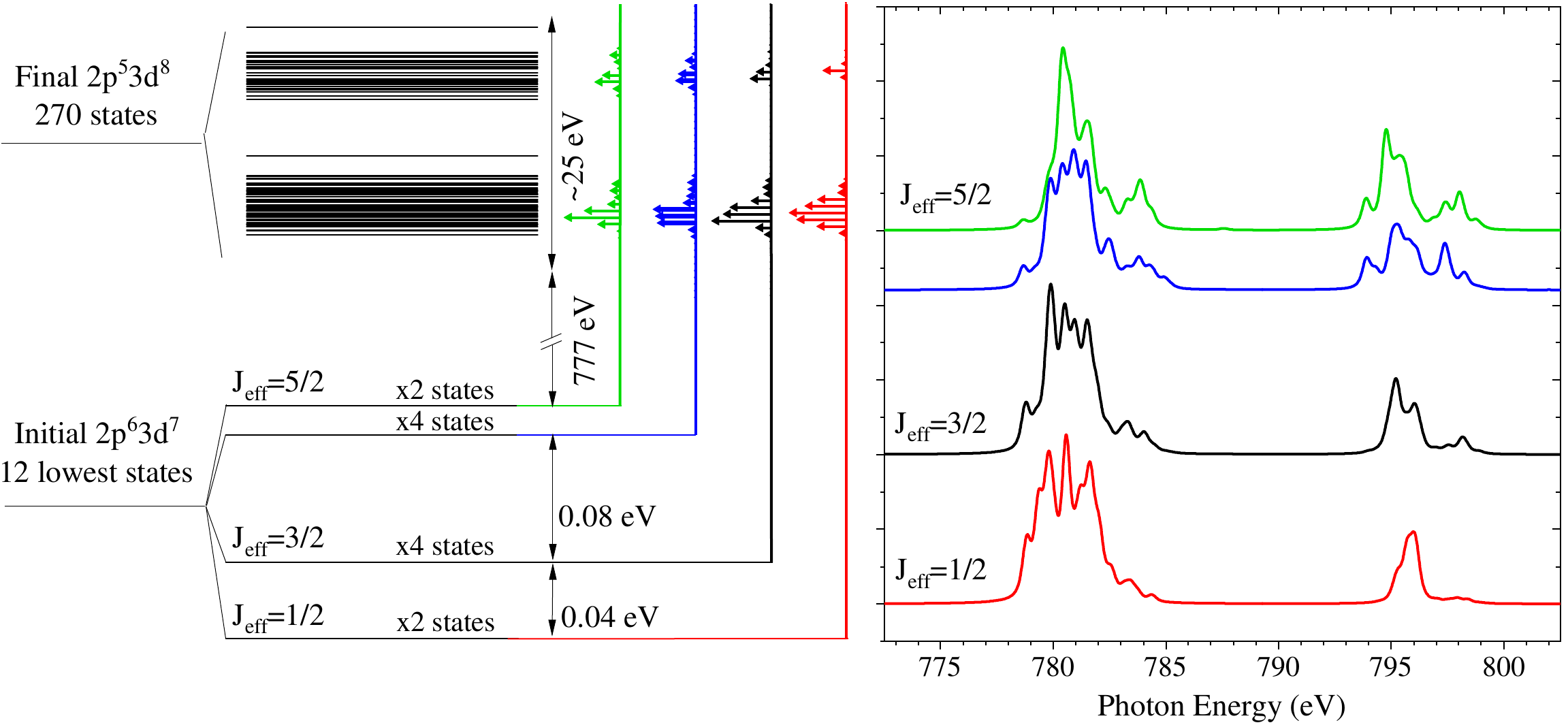}
	\caption{Left: 12 lowest states depicting the ground state and low lying excited states of a Co$^{2+}$ 3$d^7$ ion in $O_h$ crystal field. The arrows depict in an x-ray absorption experiment at the Co $L_{2,3}$ edges the allowed dipole transitions from each initial state into the 270 possible final states split by the Co $2p$ spin-orbit coupling. Right: theoretical spectra corresponding to each starting $J_\mathrm{eff}$ = $1/2$, $3/2$ and $5/2$ initial states.  }
	\label{fig:energy_level_Co2p}
\end{figure*}

In this context, XAS and XMCD at the Co $L_{2,3}$ edges combined with full-multiplet configuration-interaction cluster calculations are powerful methods to determine the ground-state quantum numbers of the Co ions as well as the relevant energy parameters like effective crystal field, SOC, and trigonal distortion \cite{Groot94,Tanaka94,Burnus2006,Chi2019,Lin2010,Bur2008}. 

Of particular interest for Co$^{2+}$ ions in octahedral coordination is that XAS and XMCD are very sensitive to which of the lowest spin-orbit split $J_\mathrm{eff}$ = $1/2$, $3/2$ and $5/2$ states is occupied.  This is illustrated in Fig.\,\ref{fig:energy_level_Co2p}: theoretical calculations for the dipole-allowed $2p^63d^7 \rightarrow 2p^53d^8$ transitions show that each of the starting $J_\mathrm{eff}$ state produces its own specific XAS spectrum. In particular, the $J_\mathrm{eff} \ = 5/2$ sextet is further split by the cubic crystal field into a quartet and a doublet. 

The splitting between the different $J_\mathrm{eff}$ states can be very small, and is definitely smaller than the experimental photon energy resolution and the inverse lifetime of the $2p$ core hole. Yet, which spectrum will be detected depends solely on which $J_\mathrm{eff}$ states are occupied. This occupation will change with temperature, leading to a non-trivial behavior of the magnetic susceptibility \cite{Bur2008}, and/or due to mixing by lower-than-octahedral crystal fields. It is interesting to note that the specificity for the $J_\mathrm{eff}$ state is not only in the linear or magnetic circular polarization dependence but already in the isotropic part of the spectrum. This makes this XAS technique particularly suitable also for polycrystalline materials.  

%%%%%%%%%%%%%%%%%%%%%%%%%%%%%%%%%%%%%%%%%%%%%%%%%%%%%%%%%%%%%%
%%%%%%%%%%%%%%%%%%%%%%%%%%%%%%%%%%%%%%%%%%%%%%%%%%%%%%%%%%%%%%
%%%%%%%%%%%%%%%%%%%%%%%%% EXPERIMENT %%%%%%%%%%%%%%%%%%%%%%%%%
%%%%%%%%%%%%%%%%%%%%%%%%%%%%%%%%%%%%%%%%%%%%%%%%%%%%%%%%%%%%%%
%%%%%%%%%%%%%%%%%%%%%%%%%%%%%%%%%%%%%%%%%%%%%%%%%%%%%%%%%%%%%%

\section{Methods}

The single crystal growth and basic characterization of BaCo$_2$(AsO$_4$)$_2$ samples used in this study can be found in Ref. \cite{Mukh2024}. The XAS and XMCD measurements were performed at the BL29 of Boreas beamline of the ALBA synchrotron radiation facility in Barcelona \cite{Bar2016}. A dark pink colored single crystal was mounted on an aluminum sample holder such that the Poynting vector and the linear horizontal (LH) polarization of the incident beam was in the $ab$ plane and the linear vertical (LV) polarization along the $c$-axis of the BaCo$_2$(AsO$_4$)$_2$ single crystal. The crystal was not oriented in plane. The experimental geometry is shown in Fig.\,\ref{fig:geometry}\,(c). The sample was cleaved \textit{in~situ} in an ultra-high vacuum chamber before being transferred into the measurement chamber. 

The beam polarization was fixed to either LH or LV during each scan, thus probing the linear dichroism (LD) between the in-plane and out-of-plane directions of the crystal, respectively. The LD measurements were conducted in zero magnetic field at temperatures $T$ = 300\,K, 200\,K, 100\,K, 50\,K and 5\,K.  For the XMCD measurements, a magnetic field of 6\,T was applied along the Poynting vector of the incident x-ray beam, which was in the $ab$ plane of the sample. The helicity of the x-ray beam was switched between left ($\sigma^{-}$) and right ($\sigma^{+}$) circular polarization for each of the energy scans. The measurements were carried out both with +\,6\,T and $-$\,6\,T applied magnetic fields to cancel any asymmetry in the background. These measurements were conducted at $T$ = 50 K and 5 K.

It should be noted that the pink colored transparent crystal of BaCo$_2$(AsO$_4$)$_2$ with an optical band gap of $\approx$ 3 eV is highly insulating. Sample charging thus prevents the widely used total electron yield (TEY) method. It is also well-known that fluorescence yield (FY) can not be used for a quantitative analysis of the XAS spectra due to distortions of the spectra as a consequence of self absorption effects. Therefore, we used the inverse partial fluorescence yield method (iPFY) developed by the Hawthorn and co-workers \cite{Ach2011,Haw2011}. They have shown that iPFY method is bulk sensitive and free of saturation effects. To measure iPFY, we used an energy sensitive silicon drift detector (SDD) to record the x-ray emission signal for a wide range of photon energies while scanning the incident X-ray photon energy through the Co $L_{2,3}$ edges.

For the simulation of the XAS and XMCD spectra, we used the Quanty code \cite{Have2012,Lu2014,Have2014,Quanty}. The crystal field splittings between the Co $e_g^\sigma$,$e_g^\pi$ and $a_{1g}$  orbitals was treated as a fitting parameter, adjusted to match the experimental spectra. The hybridization parameters and the crystal field acting on the oxygen ligands were initially derived from a tight-binding model based on the \textit{ab initio} Wannierized LDA band structure, calculated using FPLO \cite{Koep1999,fplo21}. To achieve better agreement with the experimental data, the hybridization parameters were subsequently optimized by tuning them to 84.5\% of their initial values

Field-dependent magnetization was measured using a commercial magnetometer (MPMS3, Quantum Design).

The data supporting the findings of this study are available in Edmond at https://doi.org/10.17617/3.PZLFIH, reference \cite{3.PZLFIH_2025}.

\section{Results and Discussion}

\subsection{X-ray absorption}

\begin{figure}[t]
	\centering
	\includegraphics[width=\linewidth]{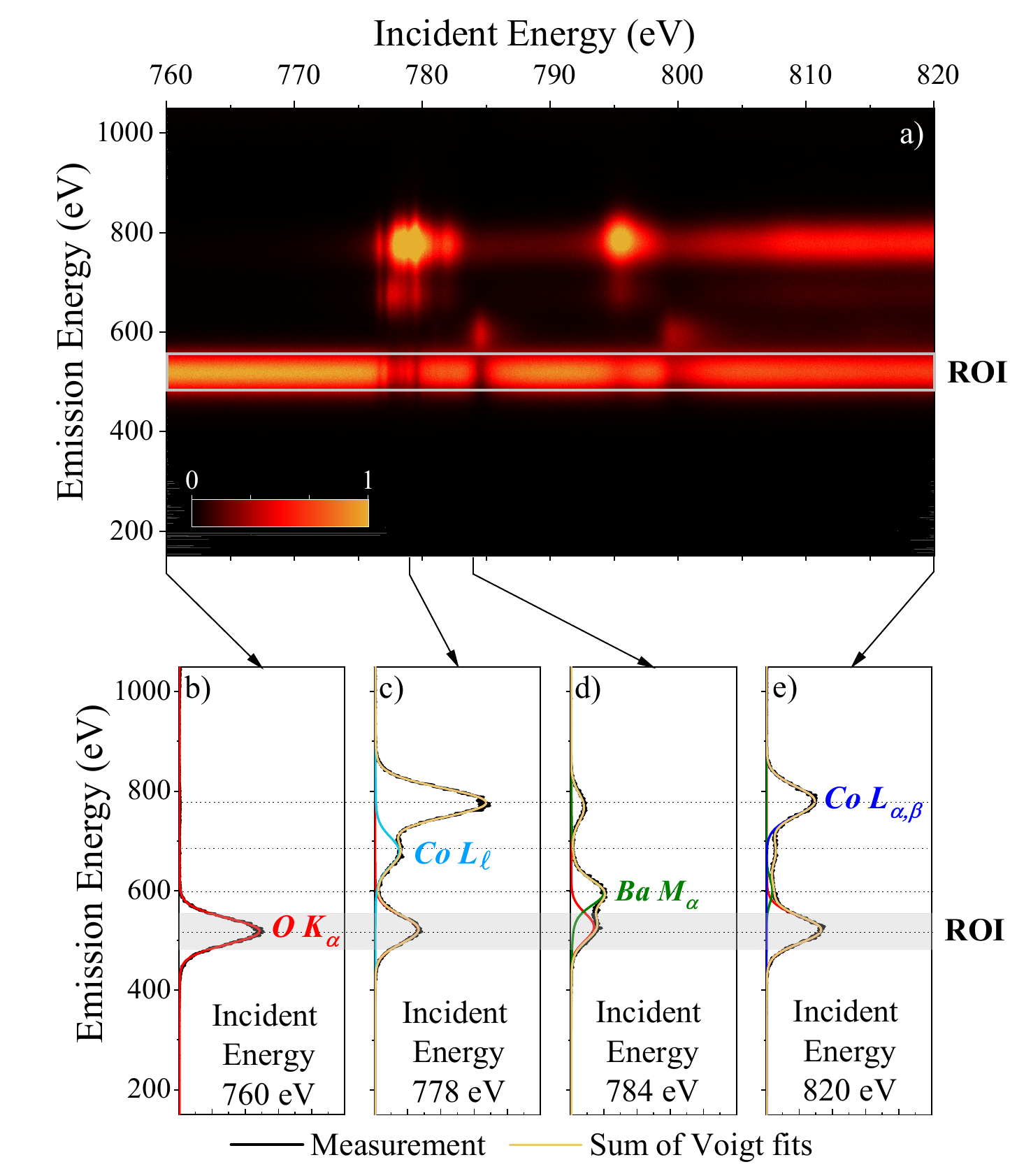}
	\caption{(a) Fluorescence yield (FY) intensity map of BaCo$_2$(AsO$_4$)$_2$ as a function of incident photon energy and emission energy, measured at 5 K using LH polarization, and normalized to the incident photon flux. The grey rectangle represents the region of interest (ROI) around the O $K_\alpha$ emission line used for obtaining the partial fluorescence yield (PFY) spectra shown in Fig.\,\ref{fig:ipfy_comp}. (b-e) Fluorescence emission spectra for selected incident photon energies corresponding to: the pre-Co and Ba edges (760 eV), at the Co $L_3$ white line (778 eV), at the Barium $M_5$ white line (784 eV), and in the post edge region (820 eV). The red, green, light blue, and dark blue lines are Voigt fits for the O $K_\alpha$, Ba $M_\alpha$, Co $L_\ell$, and Co $L_{\alpha,\beta}$ emission lines, respectively. The yellow line is the sum of the Voigt fits and black the experimental spectrum.}
	\label{fig:MCA}
\end{figure}

A typical fluorescence yield (FY) intensity map of BaCo$_2$(AsO$_4$)$_2$ measured as a function of incident photon energy and emission energy is presented in Fig.\,\ref{fig:MCA}\,(a). It can be seen that the intensities change significantly when the incident energy crosses the Co $L_{2,3}$ (778 eV, 792 eV) and Ba $M_{4,5}$ (784 eV, 798 eV) white lines. As far as emission energies are concerned, one can distinguish the O $K_\alpha$ ($\approx$ 520 eV), Ba $M_\alpha$ ($\approx$ 605 eV), Co $L_\ell$ ($\approx$ 680 eV), and Co $L_{\alpha,\beta}$ ($\approx$ 780 eV) transitions. Fig.\,\ref{fig:MCA}\,(b-e) depicts the emission spectra for selected incident photon energies corresponding to the pre-Co and Ba edges (760 eV), at the Co $L_3$ white line (778 eV), at the Barium $M_5$ white line (784 eV), and after Co and Ba white lines (820 eV).

\begin{figure}[h]
	\centering
	\includegraphics[width=\linewidth]{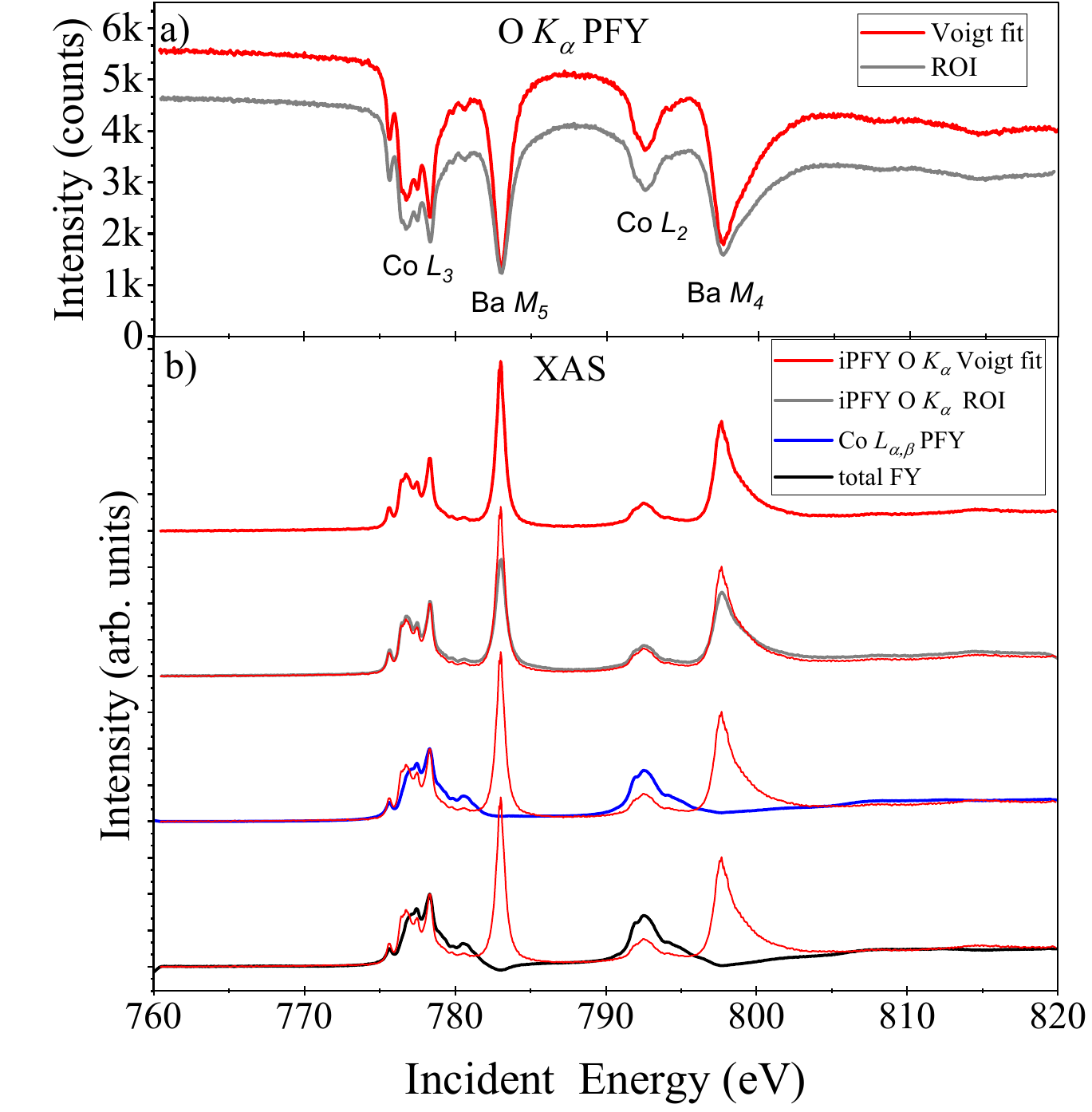}
	\caption{(a) Comparison between the O $K_\alpha$ PFY across the Co $L_{2,3}$ and Ba $M_{4,5}$ absorption edges using the grey rectangle ROI in Fig.\,\ref{fig:MCA}\,(a) and that using a Voigt fit as exemplified in Fig.\,\ref{fig:MCA}\,(b). (b) Comparison between the XAS spectra across the Co $L_{2,3}$ and Ba $M_{4,5}$ absorption edges obtained using different acquisition methods: O $K_\alpha$ iPFY using the Voigt fit (red line), O $K_\alpha$ iPFY using the ROI (grey line), Co $L_{\alpha,\beta}$ PFY (dark blue line) and total FY (black line).}
	\label{fig:ipfy_comp}
\end{figure}

Essential for the application of the iPFY method \cite{Ach2011,Haw2011} is the presence of an emission line with an energy lower than that of the absorption edge of interest. In our case we have access to the O $K_\alpha$ emission for studying the Co $L_{2,3}$ absorption spectrum of the BaCo$_2$(AsO$_4$)$_2$ material. The intensity of this O $K_\alpha$ emission descreases when the incident photon energy is tuned across the Co $L_{2,3}$ (and Ba $M_{4,5}$) absorption edges, see Fig.\,\ref{fig:ipfy_comp}\,(a). Taking an energy window (or region of interest - ROI) between 480 eV and 560 eV around the O $K_\alpha$ line as indicated by the grey box in Fig.\,\ref{fig:MCA}\,(a), we obtain the grey curve in Fig.\,\ref{fig:ipfy_comp}\,(a) as the O $K_\alpha$ PFY. The Ba $M_\alpha$ emission energy is, however, quite close to that of the O $K_\alpha$ causing a partial overlap of the emission spectra due to the limited resolution of the SDD instrument. We therefor also have extracted the O $K_\alpha$ intensity using a Voigt fit as illustrated in Fig.\,\ref{fig:MCA}\,(b-e). The resulting O $K_\alpha$ PFY is plotted as the red curve in Fig.\,\ref{fig:ipfy_comp}\,(a). The differences between the grey (ROI) and red (Voigt fit) curves are not negligible, which is of relevance for the extraction of the Co $L_{2,3}$ (and Ba $M_{4,5}$) XAS spectra.

Fig.\,\ref{fig:ipfy_comp}\,(b) shows the Co $L_{2,3}$ and Ba $M_{4,5}$ XAS spectra of BaCo$_2$(AsO$_4$)$_2$ extracted using various acquisition methods. The red curve is the iPFY \cite{Ach2011,Haw2011} spectrum based on the O $K_\alpha$ PFY signal from the Voigt fit to the O $K_\alpha$ emission line. The grey curve is the iPFY spectrum based on the ROI around the O $K_\alpha$ emission line. We can clearly observe differences in the extracted XAS spectra. Especially the strongest peaks like the Ba $M_{4,5}$ suffers from considerable loss of intensity if the Ba $M_\alpha$ emission enters the iPFY equation. The Co $L_{2,3}$ seems to be less affected, but for the remainder of our study we will utilize the O $K_\alpha$ PFY signal from the Voigt fit to ensure maximum reliability. For completeness, we show also in Fig.\,\ref{fig:ipfy_comp}\,(b) the XAS spectra obtained using the Co $L_{\alpha,\beta}$ PFY (dark blue line) and total FY (black line). Here we can clearly observe strong deviations from the O $K_\alpha$ iPFY (red line), which can be attributed, among others, to the well-known self-absorption effects in PFY and total FY for ions present in high concentrations in the material. 
\begin{figure}[h!]
	\centering
	\includegraphics[width=\linewidth]{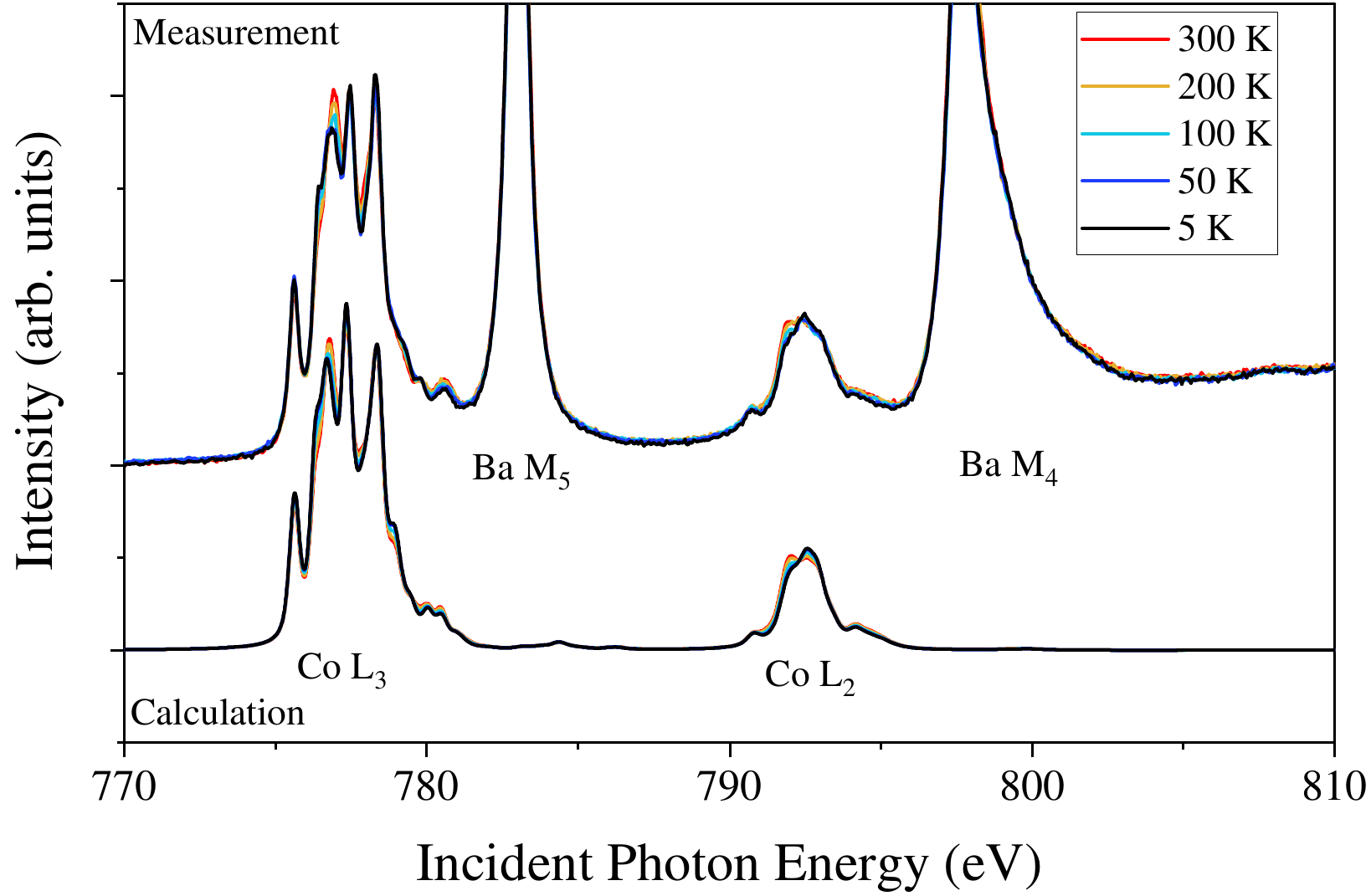}
	\caption{Comparison between experimental and calculated temperature dependence of the isotropic spectra in BaCo$_2$(AsO$_4$)$_2$.}
	\label{fig:Isospectra}
\end{figure}
In Fig.\,\ref{fig:Isospectra} we report the temperature dependence of the isotropic Co $L_{2,3}$ and Ba $M_{4,5}$ spectra. These spectra were constructed from $(2*I_\mathrm{LH} + I_\mathrm{LV})/3$ where $I_\mathrm{LH}$ and $I_\mathrm{LV}$ are the iPFY spectra taken with LH and LV polarization, respectively. The observed temperature dependence arises from the thermal population of low-lying excited states as the temperature increases. This is consistent with the expectation that the lowest-lying states for a high-spin Co$^{2+}$ ion in octahedral coordination are split in energy by the Co $3d$ SOC constant, which is of the order of 66 meV, i.e., small enough so that within a temperature range of a few hundred Kelvin, a non-trivial behavior of the magnetic susceptibility \cite{Bur2008} can be measured as well as changes in the spectra as suggested already in Fig.\,\ref{fig:energy_level_Co2p}. In fact, using parameters as will be described below, we are able to reproduce excellently the experimental data using the full-multiplet configuration-interaction calculations for a CoO$_6$ cluster. This implies that we have been successful in determining the character of the low-lying states and their energy separations. 

\subsection{Linear dichroism}\label{LD}

\begin{figure}[h!]
    \centering
    \includegraphics[width=\linewidth]{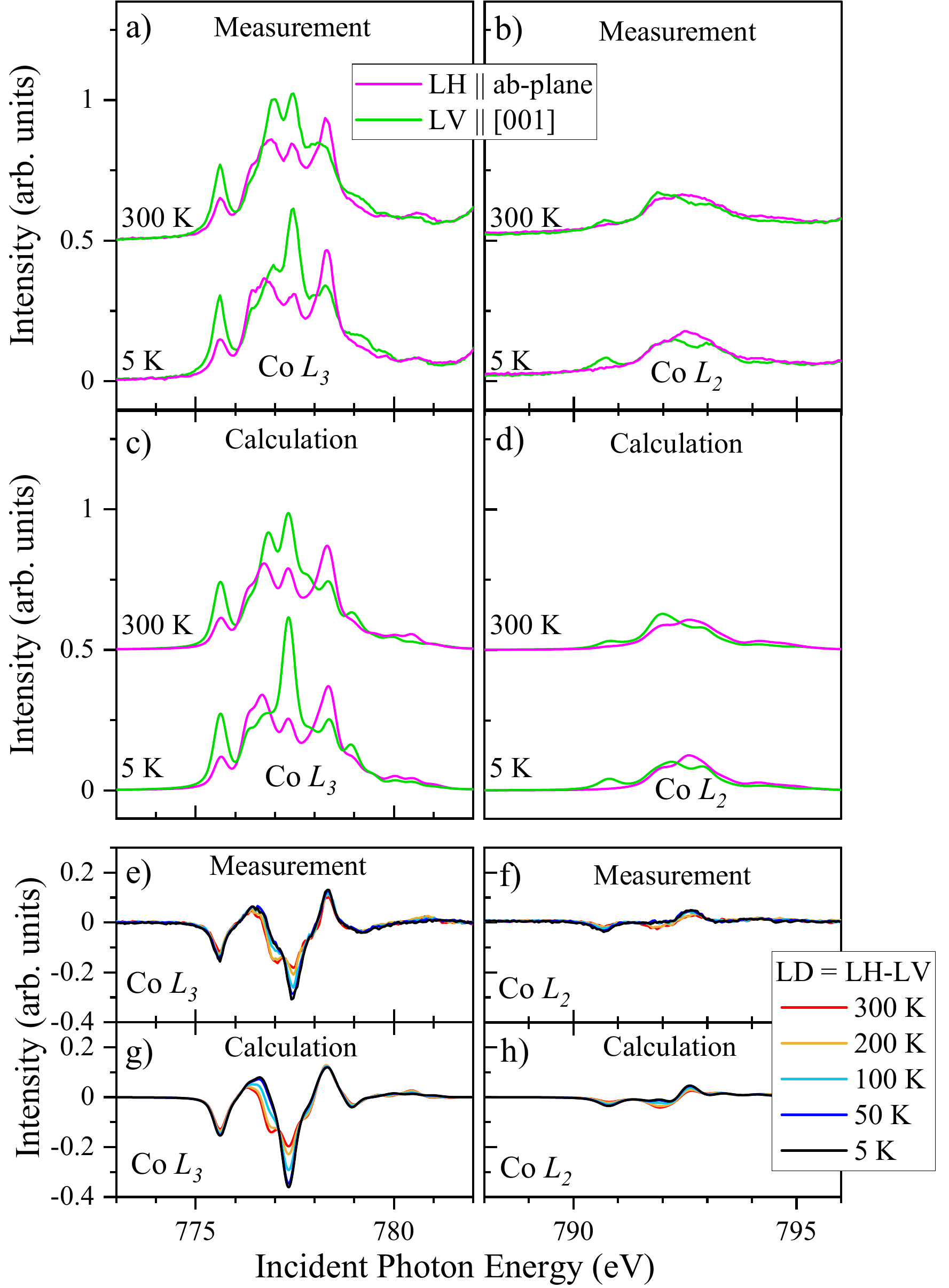}
    \caption{Top panels: polarization-dependent Co $L_{2,3}$ XAS spectra taken at 5\,K and 300\,K: (a,b) experiment and (c,d) calculation. The incident beam polarization is in the $ab$-plane (LH) and along the [001] direction (LV). Bottom panels: temperature dependence of the linear dichroic spectra $I_\mathrm{LD} = I_\mathrm{LH} - I_\mathrm{LV}$: (e,f) experiment and (g,h) calculation.}
    \label{fig:LD_T_DEP}
\end{figure}

In order to investigate whether crystal fields with lower than octahedral symmetry are present in BaCo$_2$(AsO$_4$)$_2$, we performed polarization dependent Co $L_{2,3}$ XAS measurements. Fig.\,\ref{fig:LD_T_DEP}\,(a,b) reports the experimental spectra taken at 5\,K and 300\,K with LH (in the $ab$ plane) and LV (along the [001] direction) polarization. We can observe a significant polarization dependence, which can be taken as an indication that there is a strong distortion of the CoO$_{6}$ octahedra in this material. We can also follow this polarization dependence in more detail by plotting the linear dichroism (LD) spectra defined as the difference between the LH and LV spectra. Fig.\,\ref{fig:LD_T_DEP}\,(e,b) reveals that the LD spectra become somewhat smaller in amplitude with increasing temperature but do not vanish at the highest measured temperatures.

\begin{figure}[t]
	\centering
	\includegraphics[width=\linewidth]{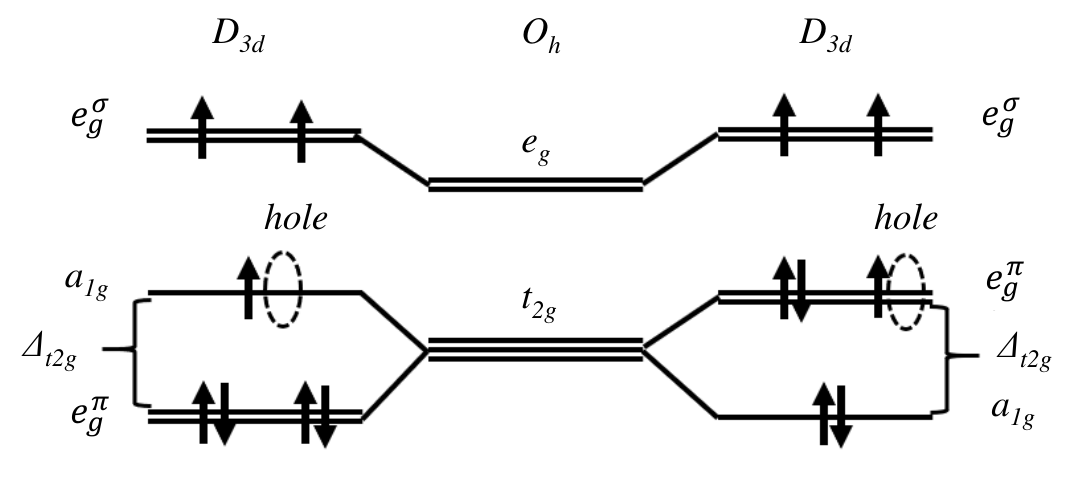}
	\caption{Schematic diagram showing the energy levels of a \textit{d} shell in $O_h$ symmetry (center) and their splitting under the influence of a trigonal distortion of $D_{3d}$ symmetry in the absence of SOC. The energy hierarchy of the orbitals under elongation along the crystallographic [001] axis is presented on the left side and that under compression is shown on the right side.}
	\label{fig:compression_elongation}
\end{figure}
  
To estimate the magnitude of the distortion of the CoO$_{6}$ octahedra we set out to simulate the spectra by introducing a trigonal distortion in our configuration-interaction cluster calculations. We limit ourselves to the $D_{3d}$ symmetry as an approximation for the Co$^{2+}$ site symmetry $C_{3}$. The $z-$axis is here defined along the [001] crystallographic axis or $c$-axis of the crystal. The energy levels of the $d$-orbitals split into a set of $a_{1g}$, $e_g^\pi$, and $e_g^\sigma$ orbitals as shown in Fig.\,\ref{fig:compression_elongation}. Depending on whether the Co$^{2+}$ octahedra undergo an elongation or compression, the $t_{2g}$ hole can reside in the $a_{1g}$ or in the $e_g^\pi$ orbitals, respectively. Making use of the orbital-polarization sum-rule in XAS \cite{csiszarControllingOrbitalMoment2005}, we can infer that the presence of an $a_{1g}$ hole will lead to a stronger integrated XAS signal when the polarization is along the z-axis, i.e., LV polarization. Here we note that the two $e_{g}^{\sigma}$ holes do not lead to a polarization dependence in the integrated XAS intensity. For an $e_{g}^{\pi}$ hole the signal will be stronger with the polarization in the ab plane, i.e., LH polarization. The experiment clearly indicates that the t2g hole is predominately in the $a_{1g}$ orbital, i.e., CoO$_6$ octahedra are elongated along the z-axis.

To be quantitative, we calculate the XAS and the LD spectra with the Quanty code \cite{Have2012,Lu2014,Have2014}. The parameters used can be found in Ref. \cite{parameters} and the results are plotted in Fig.\,\ref{fig:LD_T_DEP}\,(c,d) and in Fig.\,\ref{fig:LD_T_DEP}\,(g,h). We are able to obtain a very good agreement between calculations and measurements for all temperatures, implying that we have indeed found the crystal field and hopping parameters describing the local electronic structure of the Co$^{2+}$ ions in BaCo$_2$(AsO$_4$)$_2$.

In our calculations, we define the energy of the ionic trigonal splitting as $D^\mathrm{ion}_{trig} = E(e_g^\pi)-E(a_{1g})$; i.e., in this notation, negative values imply elongation and positive values compression of the octahedra. To arrive at the best simulation, we used a value of $D^\mathrm{ion}_{trig} = -116.8$\,meV, i.e., the octahedra are elongated and the $t_{2g}$ hole is predominately in the $a_{1g}$ orbital.

In terms of effective crystal fields, i.e., including the effect of hybridization with the O $2p$ ligands, we calculated the total energy levels by setting SOC = 0. For the effective 10$Dq^\mathrm{eff}$, we set $D^\mathrm{ion}_{trig}$=0 and find that the difference in energy between the groundstate $^{4}T_{1}$ with $t_{2g}^{5}e_{g}^{2}$ configuration and the excited state $^{4}T_{2}$ with $t_{2g}^{4}e_{g}^{3}$ configuration is 0.66\,eV; this is the value of 10$Dq^\mathrm{eff}$. For the effective trigonal distortion $D^\mathrm{eff}_{trig}$, we turn on $D^\mathrm{ion}_{trig}\,=\,-$116.8\,meV and calculate the energy of the $^{4}T_{1}$ term relative to $^{4}A$ and $^{4}E$. The energy difference is -114.08\,meV, which is thus the magnitude of $D^\mathrm{eff}_{trig}$. 

\begin{figure}[t]
	\centering
	\includegraphics[width=\linewidth]{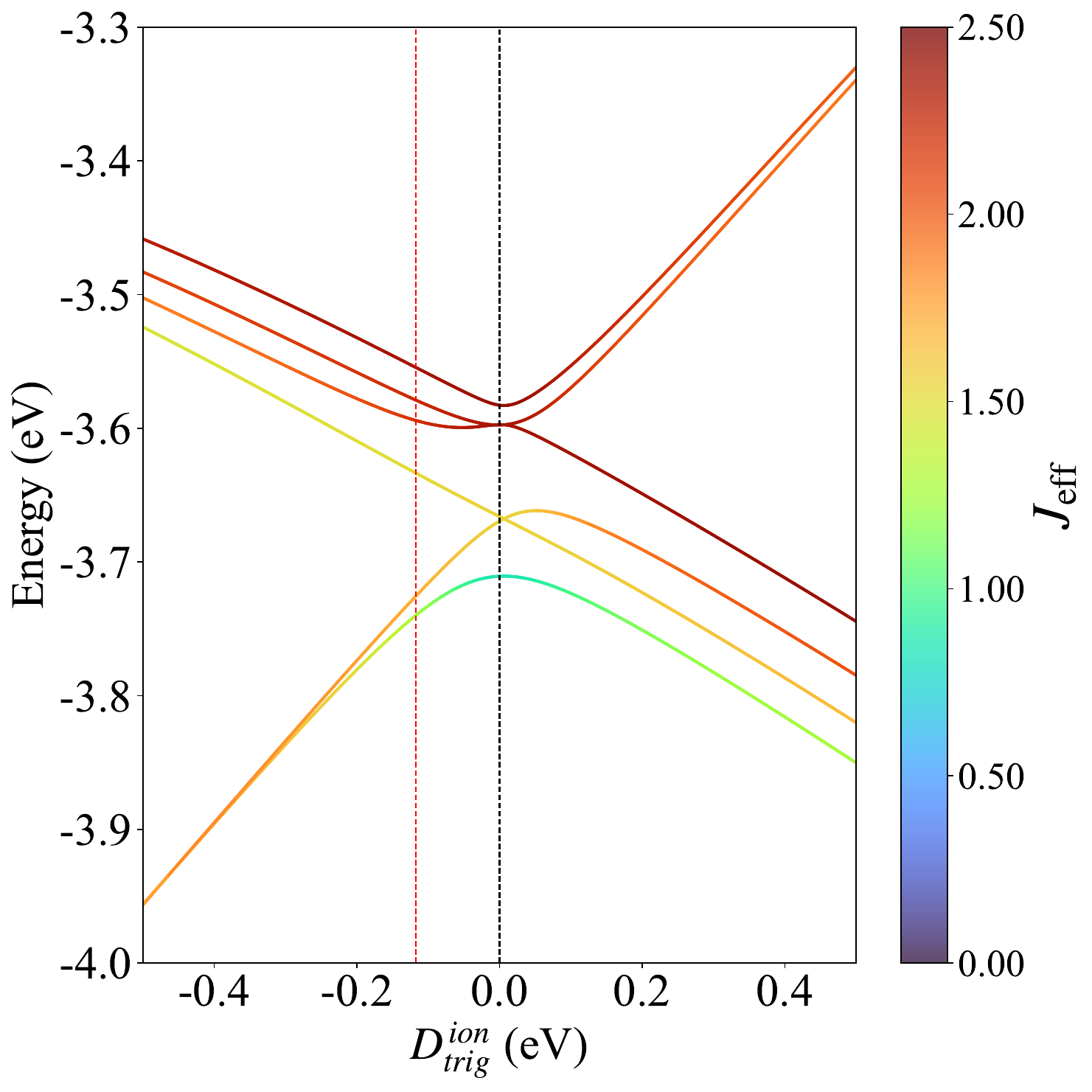}
	\caption{Energy level diagram showing the evolution of the lowest 12 eigenstates under the influence of SOC and $D_{trig}^\mathrm{ion}$. The vertical dashed black line corresponds to $D_{trig}^\mathrm{ion}$ = 0 and vertical dashed red line corresponds to $D_{trig}^\mathrm{ion}=\,-116.87$ meV. The color scale indicates the $J_\mathrm{eff}$ value.}
	\label{fig:energy_level}
\end{figure}

Since the strength of the trigonal distortion we extracted from our experiments is nearly twice the value of spin-orbit coupling constant of Co$^{2+}$ ions \cite{parameters}, it is important to visualize its impact on the $J_\mathrm{eff}$ groundstate in our material. To this end we use $\mathbf{J}_\mathrm{eff} = \mathbf{L}_\mathrm{eff} + \mathbf{S}$, where the $\mathbf{L}_\mathrm{eff}$ operator is obtained by rotating the orbital basis of the $\mathbf{L}$ operator to the cubic harmonics \cite{Agrestini2018}. The rotation matrix was modiﬁed to only keep the $t_{2g}$ subset of the $3d$ eigenorbitals. After projecting out the $e_g$ orbitals, the angular momentum operator is rotated back to the spherical harmonics. As the covalence mixes the Co $3d$ and ligand orbitals, we calculate the expectation value of the  $\mathbf{J}^2_\mathrm{eff}$ operator acting on both the Co $3d$ and ligand $d$ shell, i.e., acting on the total CoO$_6$ cluster. 

The ideal $J_\mathrm{eff}\,=\,1/2$ is valid when only the $t_{2g}$ orbitals span the Hilbert space, e.g., when the $e_g$ orbitals are completely projected out by making 10$Dq$ inﬁnitely large. A value of  $J_\mathrm{eff}$ = 0.5015 is obtained for 10$Dq$ = 10 eV, indicating that this is already close to the ideal situation. However, for the 10$Dq^\mathrm{eff}$ = 0.66\,eV value used to reproduce the experimental spectra of BaCo$_2$(AsO$_4$)$_2$, we find $J_\mathrm{eff}\,=\,0.866$. This is for the undistorted octahedron ($D^\mathrm{ion}_{trig}$=0), indicating the strong participation of the $e_g$ orbitals in addition to the $t_{2g}$. Including the trigonal distortion of $D^\mathrm{eff}_{trig}$ = -114.08\,meV found for this material, we arrive at $J_\mathrm{eff}\,=\,1.302$, a value substantially far away from the ideal case. Nevertheless, this state is still doubly degenerate. In Fig.\,\ref{fig:energy_level} we show the evolution of the energies of the lowest 12 eigenstates as a function of $D^\mathrm{ion}_{trig}$ together with the colors indicating the $J_\mathrm{eff}$ values.

\subsection{X-ray magnetic circular dichroism}

\begin{figure}[t]
    \centering
    \includegraphics[width=\linewidth]{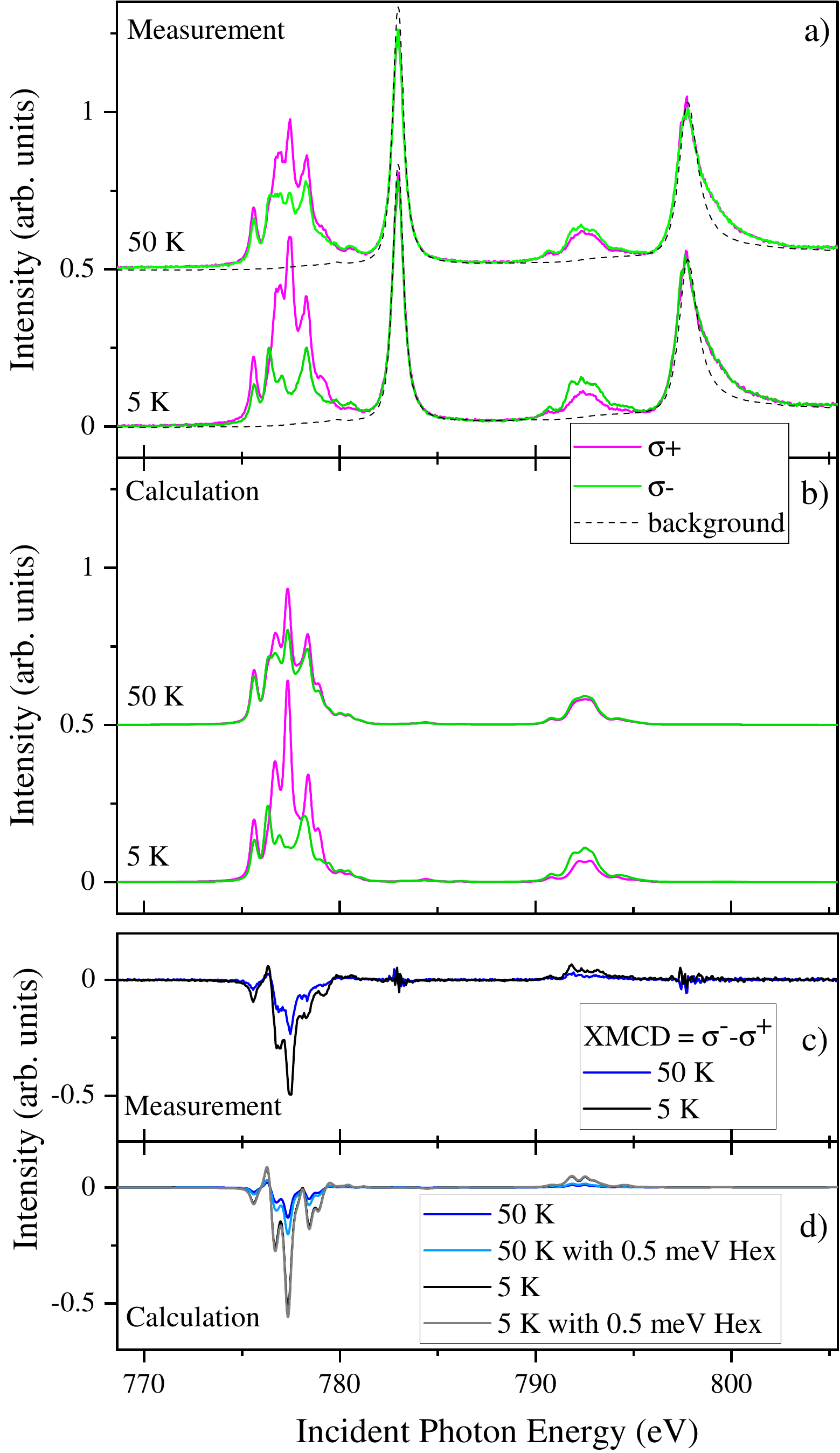}
    \caption{(a) Experimental and (b) calculated Co $L_{2,3}$ XAS of BaCo$_2$(AsO$_4$)$_2$ at 5 K and 50 K for positive ($\sigma^+$) and negative ($\sigma^-$) helicity of the circularly polarized incident beam for an applied field of \textit{B} = 6 T along the $ab-$plane of the sample. (c) Experimental and (d) calculated XMCD signal obtained from the XAS spectra shown in (a) and (b), respectively (see text for details).}
    \label{fig:xmcd}
\end{figure}

To experimentally determine the spin moment $2\braket{S_x}$ and orbital moment $\braket{L_x}$ of BaCo$_2$(AsO$_4$)$_2$, we performed XAS measurements at 5\,K and 50\,K by applying an in-plane magnetic field of 6\,T. BaCo$_2$(AsO$_4$)$_2$ displays a complex magnetic phase diagram for magnetic fields below 1\,T \cite{Zho2020,Mukh2024}. Whereas, under a magnetic field of 6\,T, the magnetic moments in the sample are fully aligned in the direction of the applied magnetic field, i.e., in the $ab$-plane (see Fig. \ref{fig:enter-label} in Appendix A). In Fig. \ref{fig:xmcd}(a) the XAS measurements performed using $\sigma^+$ and $\sigma^-$ polarization are presented. The XMCD spectra were obtained by taking the difference $(\sigma^--\sigma^+)$, which is shown in Fig. \ref{fig:xmcd}\,(c). 

The expectation values $2\braket{S_x}$ and $\braket{L_x}$ can be obtained by application of XMCD sum rules \cite{Thole1992,Thole1993}
as given in Eq.\ref{sumrules1} and Eq.\ref{sumrules2}. 

\begin{equation}
    L = \frac{4}{3}\frac{\int_{L_{2,3}}(\sigma^+-\sigma^-)dE}{\int_{L_{2,3}}(\sigma^++\sigma^-)dE}
    \label{sumrules1}
\end{equation}
\begin{equation}
    2S + 7T =2\frac{\int_{L_{3}}(\sigma^+-\sigma^-)dE-2\int_{L_{2}}(\sigma^+-\sigma^-)dE}{\int_{L_{2,3}}(\sigma^++\sigma^-)dE}N_h
    \label{sumrules2}
\end{equation}

In order to determine the relevant sum-rule integrals from the spectra, we removed the step-edge and Ba peak contribution by subtracting the background as shown by the dashed line in Fig. \ref{fig:xmcd}\,(a). The spectra were also normalized to the area of the isotropic spectrum, i.e., $(\sigma^++\sigma^-)/2$ of the $L_3$ edge. The magnetic dipole moment $T_x$ is typically a small number (0.012 for the groundstate in our cluster calculation) compared to $2\braket{S_x}$, and hence will be neglected. Applying the sum rules to the 5 K spectra and using $N_h$=2.89 from the cluster calculation yields $2\braket{S_x} = -1.99$ and $\braket{L_x} = -0.78$.

To obtain more reliable results, we first calculated the XMCD spectra, see Fig. \ref{fig:xmcd}\,(b), using the same parameters as for the LD analysis \cite{parameters}. It can be seen that the calculated spectra match quite well with the experimental ones. We then extract the desired quantum numbers from the ground state of the cluster. We find $2\braket{S_x}\,=-2.014$ and $\braket{L_x} = -0.710$. The difference is less than 10\%, indicating that the sum rules work rather well for this compound. This close agreement, indicates that the subtraction of the Ba contribution from the XAS spectra was effective. The total magnetic moment projected on the applied magnetic field direction, $i.e.$,  $M_x=|\braket{L_x}+2\braket{S_x})|$, is 2.72 $\mu_B/$Co and \sr{2.82 $\mu_B/$CoO$_{6}$ cluster}. These values are in reasonable agreement with the saturation magnetization of 2.92\,$\mu_B/$Co from the $M(H)$ in-plane measurements presented in Fig. \ref{fig:enter-label}. The effective $g$-factors obtained from the relation $M=g M_{J_\mathrm{eff}}$ with $M_{J_\mathrm{eff}}=1/2$ from our cluster calculations using an applied field of 6 T are: $g_{ab}$= 5.64 and  $g_{c}$= 2.49. These values compare well with $g_{ab}$ = 5.0 and $g_{c}$ = 2.5-2.7 from previously reported experiments \cite{Reg2018,Hol2023}. The significant difference between $g_{ab}$ and $g_{c}$ contributes to the strong magnetocrystalline anisotropy observed in this material.
Concerning the 50 K data, we notice that the simulation produces a smaller XMCD effect than the experiment. This can be resolved by adding an exchange field of 0.5 meV to the 6 T magetic field, as to represent short-ranged inter-Co exchange interactions.

\subsection{Discussion}

Using polarization dependent XAS, we have experimentally determined the energy of the trigonal splitting $D_\mathrm{trig}^\mathrm{eff}\,=\,-114.08\,$meV in BaCo$_{2}$(AsO$_{4})_{2}$. \sr{The value of $|D_\mathrm{trig}^\mathrm{eff}|$ found here is in good agreement with that reported in recent Raman spectroscopy studies \cite{Mou2024}.  Note that the negative sign of  $D_\mathrm{trig}^\mathrm{eff}$ in our studies is only due to our notation $D^{ion}_{trig} = E(e_g^\pi)-E(a_{1g})$, which is the opposite of what is reported in Ref. \cite{Mou2024}.} This value is nearly twice the value of atomic spin-orbit coupling constant of 66\,meV of Co$^{2+}$ ions. This implies significant deviations from the ideal $J_\mathrm{eff}\,=\,1/2$ groundstate desired for the formation of a pure Kitaev quantum spin liquid. The magnitude of trigonal distortion obtained here is consistent with \textit{ab~initio} density-functional theory, which takes into account realistic crystal-field distortions \cite{Das2021}. \textit{Ab~initio} calculations of effective spin models for BaCo$_{2}$(AsO$_{4})_{2}$ yield a negligible Kitaev term \cite{Maks2022}, whereas experimental estimates of the Kitaev term vary both in sign and magnitude \cite{Lee2024,Lee2025,Safari2024,Maksimov2025,Devillez2025}. \sr{A time-domain terahertz spectroscopy study reported a broad magnetic continuum in another part of the phase diagram, i.e., in a magnetic field of 4\,T applied in the out-of-plane direction. The magnetic continuum is found to be suppressed below $T_\mathrm{N}$, and this behavior is considered to be compatible with the spin-liquid models describing fractionalized excitation \cite{Zha2022}.}
 
\sr{On the other hand, application of hydrostatic pressure changes relative stability of the different phases in the field-temperature phase diagram, but it does not suppress magnetic order in any prominent way \cite{Huy2022,Mukh2024}.} The elongative trigonal distortion of CoO$_6$ observed in our experiments suggests that uniaxial compressive strain applied along the $c$-axis or a stretching of $ab$-plane might be a more effective method to suppress the conventional long-range magnetic order in BaCo$_{2}$(AsO$_{4})_{2}$, although c-axis pressure initially increases $T_N$ \cite{Mukh2024}. A suppression of $T_N$ through strain engineering of another honeycomb lattice Cu$_{3}$Co$_{2}$SbO$_{6}$ in thin films form has been reported \cite{Kim2024}. 

%%%%%%%%%%%%%%%%%%%%%%%%%%%%%%%%%%%%%%%%%%%%%%%%%%%%%%%%%%%%%%
%%%%%%%%%%%%%%%%%%%%%%%%%%%%%%%%%%%%%%%%%%%%%%%%%%%%%%%%%%%%%%
%%%%%%%%%%%%%%%%%%%%%%%%% CONCLUSION %%%%%%%%%%%%%%%%%%%%%%%%%
%%%%%%%%%%%%%%%%%%%%%%%%%%%%%%%%%%%%%%%%%%%%%%%%%%%%%%%%%%%%%%
%%%%%%%%%%%%%%%%%%%%%%%%%%%%%%%%%%%%%%%%%%%%%%%%%%%%%%%%%%%%%%

\section{Conclusions}
We have successively carried out polarization dependent soft x-ray absorption spectroscopy on the Kitaev candidate honeycomb lattice compound BaCo$_{2}$(AsO$_{4})_{2}$ using the iPFY method, which is the most reliable manner to collect spectra suitable for quantitative analysis on highly insulating samples. Our results indicate a sizable trigonal distortion causing a splitting of $\approx -114.08$\,meV in this compound. The observation of strong temperature dependence of LD confirms the thermal population of excited states in the temperature regime above 100\,K. From the XMCD sum rules and cluster-model analysis of the spectra, we obtain the total in-plane magnetic moment of 2.72-2.77 $\mu_B/$Co which is in reasonable agreement with the saturation value of in-plane magnetic moment obtained from the magnetization measurement.
Our findings suggest that the application of an uni-axial strain is a possible route to possibly tune the system into a quantum spin liquid regime.

%%%%%%%%%%%%%%%%%%%%%%%%%%%%%%%%%%%%%%%%%%%%%%%%%%%%%%%%%%%%%%
%%%%%%%%%%%%%%%%%%%%%%%%%%%%%%%%%%%%%%%%%%%%%%%%%%%%%%%%%%%%%%
%%%%%%%%%%%%%%%%%%%%%% ACKNOWLEDGMENTS %%%%%%%%%%%%%%%%%%%%%%%
%%%%%%%%%%%%%%%%%%%%%%%%%%%%%%%%%%%%%%%%%%%%%%%%%%%%%%%%%%%%%%
%%%%%%%%%%%%%%%%%%%%%%%%%%%%%%%%%%%%%%%%%%%%%%%%%%%%%%%%%%%%%%

\begin{acknowledgments}

M.M.F.-C. and S. R. thank Stefano Agrestini and Javier Herrero-Martín for valuable discussions. 
The work at the MPI-CPfS Dresden was partially supported by SFB1143 (project number 247310070).  
M.M.F.-C. greatly acknowledges funding from the  Deutsche Forschungsgemeinschaft (DFG, German Research Foundation) Grant No. 387555779. The work in Leipzig and Augsburg was funded by the Deutsche Forschungsgemeinschaft (DFG, German Research Foundation) -- TRR 360 -- 492547816 (subproject B1). P.K.M. acknowledges the financial support of Alexander von Humboldt Foundation. XLD and XMCD measurements were performed at ALBA under Proposal No. ID 2023097766. 

\end{acknowledgments}

\appendix
\section{}

\begin{figure}[h]
    \centering
    \includegraphics[width=\linewidth]{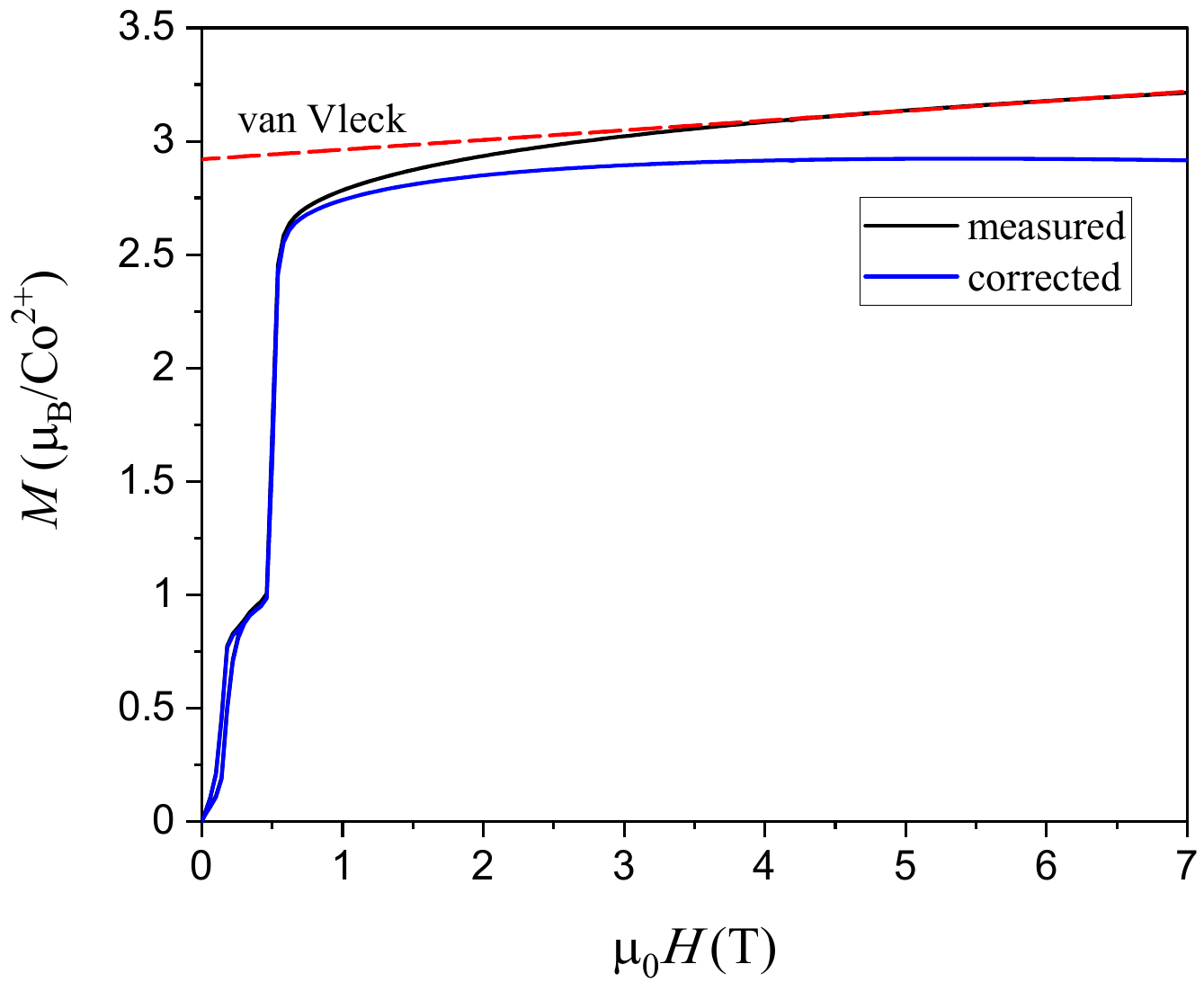}
    \caption{Magnetization $M(H)$ measured at $T\,=\,2\,$K with the applied magnetic field parallel to the $ab-$plane. The dashed red line represents the van Vleck contribution. Legends measured shows as-measured $M(H)$ curve, and  corrected represents the $M(H)$ data after correcting for the van Vleck contribution.}
    \label{fig:enter-label}
\end{figure}

In order to compare the total in-plane magnetic moment $M_x$ obtained from the XMCD sum rules with that from the magnetization measurements, we performed $M(H)$ with the magnetic field $H$ applied parallel to the $ab$-plane. The measurements were performed at $T=\,$2\,K. In zero applied magnetic field, BaCo$_{2}$(AsO$_{4})_{2}$ orders antiferromagnetically below $T_N\,\approx$ 5.4\,K into a double zig-zag, incommensurate helical spin structure \cite{Reg1977,Reg2018}. Upon application of in-plane magnetic field, the compound goes through a series of complex phases \cite{Zho2020,Mukh2024}, and finally reaching the full spin polarized state of the Co$^{2+}$ ions in the applied magnetic field $\mu_{0}H\,\approx\,$2\,T as can be seen from Fig.\ref{fig:enter-label}. Above 2\,T, the $M(H)$ increases due to the the van Vleck paramagnetic contribution, which is marked by the red dashed line in  Fig. \ref{fig:enter-label}. After subtracting the van Vleck contribution, we obtain a saturation in-plane magnetic moment of 2.92\,$\mu_B/$Co.
\balance
\bibliographystyle{unsrt}

\end{document}